# Using Optimization to Obtain a Width-Independent, Parallel, Simpler, and Faster Positive SDP Solver[*]

Zeyuan Allen-Zhu
zeyuan@csail.mit.edu
Princeton University

Yin Tat Lee
yintat@mit.edu
MIT

Lorenzo Orecchia
orecchia@bu.edu
Boston University

January 11, 2016

## Abstract

We study the design of polylogarithmic depth algorithms for approximately solving packing and covering semidefinite programs (or positive SDPs for short). This is a natural SDP generalization of the well-studied positive LP problem.

Although positive LPs can be solved in polylogarithmic depth while using only $\widetilde{O}(\log^2 n/\varepsilon^2)$ parallelizable iterations [4, 33], the best known positive SDP solvers due to Jain and Yao [18] require $O(\log^{14} n/\varepsilon^{13})$ parallelizable iterations. Several alternative solvers have been proposed to reduce the exponents in the number of iterations [19, 30]. However, the correctness of the convergence analyses in these works has been called into question [30], as they both rely on algebraic monotonicity properties that do not generalize to matrix algebra.

In this paper, we propose a very simple algorithm based on the optimization framework proposed in [4] for LP solvers. Our algorithm only needs $\widetilde{O}(\log^2 n/\varepsilon^2)$ iterations, matching that of the best LP solver. To surmount the obstacles encountered by previous approaches, our analysis requires a new matrix inequality that extends Lieb-Thirring's inequality, and a sign-consistent, randomized variant of the gradient truncation technique proposed in [3, 4].

---

# 1 Introduction

Solvers for linear programs (LPs) and semidefinite programs (SDPs) are important algorithmic tools for many computational tasks, spanning the fields of computer science, operations research, statistics, and applied mathematics. Although polynomial-time generic solvers for LPs and SDPs have been known for a long time, their performance is often unsatisfactory in the big-data scenario.

In the past two decades, a significant amount of attention has been paid towards a special class of LPs and SDPs, known as positive LPs [23] and positive SDPs [20] respectively. At a high level, positive LPs are characterized by non-negative variables and a non-negative constraint matrix; similarly, positive SDPs are described by positive semidefinite (PSD) matrix variables and a family of PSD matrices as constraints. In this paper, we are interested in solving positive SDPs, formally defined as follows.

**Positive SDP.** Given $m \times m$ PSD matrices $A_1, A_2, \ldots, A_n$, positive SDP (after putting in its standard form) refers to the following pair of SDPs:[1]

$$\text{Packing SDP:} \quad \max_{x \geq 0} \left\{ \mathbb{1}^T x \,:\, \sum_{i=1}^n x_i A_i \preceq I \right\} , \tag{1.1}$$

$$\text{Covering SDP:} \quad \min_{Y \succeq 0} \left\{ \operatorname{Tr}(Y) \,:\, A_i \bullet Y \geq 1 \ \forall i \in [n] \right\} . \tag{1.2}$$

Since the two programs are dual to each other, let us denote by OPT the optimal value to both of them. Also, let $x^*$ be any optimal solution for the packing SDP (1.1). We say that $x \geq 0$ is a $(1-\varepsilon)$-approximation to the packing SDP if $\sum_{i=1}^n x_i A_i \preceq I$ and $\mathbb{1}^T x \geq (1-\varepsilon)\mathsf{OPT}$, and $Y \succeq 0$ a $(1+\varepsilon)$-approximation to the covering SDP if $A_i \bullet Y \geq 1$ for all $i \in [n]$ and $\operatorname{Tr}(Y) \leq (1+\varepsilon)\mathsf{OPT}$.

In this paper, we assume without loss of generality that

$$\min_{i \in [n]}\{\|A_i\|_{\mathsf{spe}}\} = 1 \quad \text{where } \|A_i\|_{\mathsf{spe}} \text{ is the spectral norm of } A_i \ ,$$

since otherwise one can scale all $A_i$ by a constant factor, and the solution OPT as well as $x^*$ are only affected by this same constant factor. We denote by $\mathbf{A} = (A_1, \ldots, A_n)$.

**History.** Positive SDP instances have been used to model a large numer of computational problems, such as MAX-CUT [14, 20], sparse PCA [14], coloring [14], the ARV relaxation of SPARSESTCUT [13] and BALANCEDSEPARATOR [7, 28], and many others. Positive SDPs also found application in computational complexity, where they were crucial in establish the QIP = PSPACE equivalence [15], as well as in quantum interactive proofs [16] and quantum zero-sum games [17]. In addition, techniques developed in this line of research have also inspired many other important results, most notably regarding spectral graph theory [2, 27, 28].

While there has been a lot of research on the fast approximate solution of positive LPs [3, 4, 6, 8–12, 21, 23–26, 32, 36, 37], the more general positive SDP case has lagged somewhat behind. Most known positive SDP solvers [5, 7, 13–17] demand a parallel running time that is $\mathsf{polylog}(nm/\varepsilon) \cdot \mathsf{poly}(\rho)$ in order to produce a $(1 \pm \varepsilon)$ approximation of the optimal value. In this expression, $\rho$ is a "width" parameter that depends on the *numeric value* of the SDP and that can sometimes be as large as $\mathsf{poly}(n, m)$.

---

[1]The most general form of covering SDP can be written as follows. Given $m \times m$ PSD matrices $C, A_1, \ldots, A_n$, and non-negative scalars $b_1, \ldots, b_n$, a general covering SDP is to

minimize $C \bullet Y$ subject to the constraint that $A_i \bullet Y \geq b_i$ for each $i \in [m]$ and $Y \succeq 0$.

It is a simple exercise, but anyways proved in [30, Appendix A], to see that the above general form can be easily translated into our standard form. This is also true for packing SDP.

| Problem | Paper | Parallel Depth Per Iteration | Number of Iterations |
|---------|-------|------------------------------|----------------------|
| p/c LP  | [23]  | $\log(nm)$                   | $\log^2(nm)/\varepsilon^4$ |
| p/c LP  | [4]   | $\log(nm)$                   | $\log^2(nm)/\varepsilon^3$ |
| p/c SDP | [18]  | $\mathsf{polylog}(nm) \cdot \mathsf{poly}(1/\varepsilon)$ | $\log^{14}(nm)/\varepsilon^{13}$ |
| p/c SDP | [19, 30] | $\log^2(nm)/\varepsilon$  | $\log^2(nm)/\epsilon^4$, in doubt [a] |
| p/c SDP | **[this paper]** | $\log^2(nm)/\varepsilon$ | $\log^2(nm)/\varepsilon^3$ [b] |

Table 1: Comparisons of asymptotic running times among width-independent approximate solvers for positive LPs and SDPs. Notice that each iteration of a SDP solver requires a $1/\varepsilon$-dependance to approximate the matrix exponential using the Johnson-Lindestrauss Lemma [30].

[a] See Section 2 for details.
[b] The present paper only discusses the algorithm that converges in $\log^2(nm)/\varepsilon^3$ iterations. It can be improved to $\log^2(nm)\log(1/\varepsilon)/\varepsilon^2$ for positive SDP using exactly the same technique provided by Wang et al. [33] for positive LP. We shall include a detailed proof of this in a next version of this paper.

In a seminal work in 1993, Luby and Nisan [23] introduced the first width-independent and polylogarithmic-parallel-time positive LP solver. Based on this breakthrough, in 2011, Jain and Yao [18] proposed the first approximate positive-SDP solver that is *width-independent* and whose parallel running time is only $\mathsf{poly}(\log n, \frac{1}{\varepsilon})$. In fact, their algorithm is a faithful generalization of the positive LP solver of Luby and Nisan [23] to positive SDPs. Although the convergence rate (i.e., number of parallelizable iterations) required by Luby and Nisan's algorithm is only $O(\log^2(nm)/\varepsilon^4)$, the convergence rate of Jain and Yao's is as large as $O(\log^{14}(nm)/\varepsilon^{13})$ (see Table 1). This significant loss in the running time stems from the harder task of computing with matrices and in particular by the loss of commutativity in matrix algebra with respect to the vector setting.

The poor theoretical performance of [18] has attracted some researchers to study alternative positive-SDP solvers. Motivated by Young's algorithm [36] for positive LPs, two alternative solvers have been proposed [19, 30]. However, the theoretical convergence of these two new solvers remains unclear, as the correctness of both convergence analyses has been called into question. The issue with the algorithm of [30] is explicitly stated in the latest ArXiv version of that paper [31]. A similar issue has been identified [29, 35] with the proof of [19]. In a nutshell, the proof difficulties in both works arise because Young's algorithm, in its current form, relies on a monotonicity argument. While such monotonicity holds naturally in the vector (i.e., LP) case, it does not generalize to the matrix (i.e. SDP) world. See Section 2 for a detailed discussion of this.

As a result, the best parallel running time of width-independent positive SDP solvers remains to be $O(\log^{14}(nm)/\varepsilon^{13})$ due to Jain and Yao [18].

**This Paper.** In this paper, we present an algorithm `PosSDPSolver`$(\mathbf{A}, \varepsilon)$ that runs only in $O(\frac{\log n \cdot \log(nm/\varepsilon)}{\varepsilon^3})$ iterations. This matches the best convergence rate of the width-independent parallel positive LP solver [4], and is a significant improvement over the best known width-independent positive SDP solver by Jain and Yao [18]. It is also an improvement over the solvers of [30] and [19], even if their analyses can be fixed. (See Table 1.)

Our algorithm is also much simpler than all the previous width-independent positive SDP solvers, as it avoids the use of "phases" and restarts that are required by previous solvers [18, 19, 30]. Our algorithm is simply divided into $O(\frac{\log n \cdot \log(nm/\varepsilon)}{\varepsilon^3})$ iterations. Starting from some initial vector $x \geq 0$, in each iteration, we compute $n$ matrix exponential computations $A_1 \bullet e^\Psi, \ldots A_n \bullet e^\Psi$ in parallel for some symmetric matrix $\Psi$ satisfying $\|\Psi\|_{\mathsf{spe}} \leq O(\log(nm)/\varepsilon)$, and then change $x_i$ according to the value of $A_i \bullet e^\Psi$. This same algorithm simultaneously produces $1 \pm O(\varepsilon)$ approximate solutions to the packing SDP (1.1) and the covering SDP (1.2),.

We remark here that, as originally put forward by Arora and Kale [7], and then formally established by Peng and Tangwongsan [30], each of our iterations can be implemented to run in $O(\log^2(nm)/\varepsilon)$ parallel time after some simple preprocessing. In fact, such computations are required by all the previous width-independent positive SDP solvers.

**Our Techniques.** Our algorithm is directly based on the optimization framework of the positive LP solver recently put forward by Allen-Zhu and Orecchia [4]. The non-commutativity introduced by matrices creates significant obstascles and technical challenges that have forced us to make both our algorithm and analysis different from [4].

To begin with, just like the result in [4], we interpret the positive SDP problem as a purely optimization question, i.e., to minimize $f(x)$ for some convex function $f = f^{sdp}$ that is an SDP extension over its LP choice $f^{lp}$ proposed in [4]. In each iteration of our algorithm, we compute the coordinate gradient $\nabla_i f(x) \stackrel{\text{def}}{=} A_i \bullet e^{\frac{1}{\mu}(\sum_{i \in [n]} x_i A_i - I)} - 1$ for each $i \in [n]$.

AN OLD STORY. In [4], the authors update each $x_i$ as follows. They first define the truncated gradient by letting $\xi_i$ be essentially $\min\{1, \nabla_i f(x)\}$.[2] Next, update each $x_i \leftarrow x_i \cdot e^{-\alpha \xi_i}$ for some global parameter $\alpha = \Theta(\varepsilon^2/\log(nm)) > 0$.

The key idea behind the convergence result of [4] is that, if one changes $x$ according to the rule above, then for each "important" $i \in [n]$ (i.e., coordinates $i$ satisfying $\nabla_i f(x) \notin [-\varepsilon, \varepsilon]$), we have that $\nabla_i f(x)$ is guaranteed to change multiplicatively within a factor of $1 \pm \frac{1}{2}$ as $x$ changes, and therefore the sign of $\nabla_i f(x)$ for each important $i$ remains the same before and after each update. This leads to the conclusion that the objective value $f(x)$ effectively decreases during each iteration.

Unfortunately, this "multiplicative-change" guarantee, which is a crucial component of most width-independent solvers, is false in the SDP setting.

OUR NEW IDEAS. In this paper, we make two important observations. First, suppose for a moment that $x$ is updated in a sign-consistent manner: either it non-decreases or it non-increases for all the coordinates. Even under this sign-consistent assumption, $\nabla_i f(x)$ does not necessarily remain of the same sign for each important coordinate $i$, so the previous analysis of [4] still fails in the SDP setting. However, under this sign-consistencty assumption, we can show that a carefully chosen weighted summation of $\nabla_i f(x)$ does maitain the same sign. This consideration is sufficient to prove that the objective signficantly decreases at every iteration. To show that the weighted summation remains of the same sign, we require a generalization of the Lieb-Thirring inequality. To the best of our knowledge, this is a new matrix inequality, which may be of independent interest. We shall discuss the relation between our generalizaiton of Lieb-Thirring and positive SDPs in Section 2.

Finally, to ensure that $x$ is updated in a sign-consistent manner, we introduce randomness as follows. We flip an unbiased coin at each of our iterations, and choose to either update $x_i$'s in a non-decreasing manner (therefore ignoring all coordinates $i$ with $\nabla_i f(x) > 0$), or in a non-increasing manner (therefore ignoring all coordinates $i$ with $\nabla_i f(x) < 0$). Such a random choice can be shown to decrease the objective $f(x)$ well *in expectation*, but adds a lot difficulty to the analysis of the covering SDP. In short, after such randomness is introduced, the old analysis of [4] only gives a solution $Y$ whose expectation $\mathbf{E}[Y]$ is feasible to the covering SDP (1.2): that is, $A_i \bullet \mathbf{E}[Y] \leq 1$ for each $i \in [n]$. Such a result is totally useless because we need $A_i \bullet Y \leq 1$ for each $i \in [n]$, and therefore we need to propose a totally different analysis that bypasses this difficulty (see Section 6).

**Conclusion.** In this paper we show that the positive LP solver by Allen-Zhu and Orecchia [4] can be extended to the SDP setting without any asymptotic loss in the convergence rate.

---
[2]There is an optimization insight behind why such a truncation is needed and we refer the interested readers to the introduction of [4].

At a high level, to convert *any* positive LP solver to SDP, one needs to tradeoff between (a) "what is allowed to be changed in the algorithm without hurting its performance" and (b) "what must be changed in order to work with matrix algebra". In this paper, we make use of the optimization framework of [3], which gives us the greatest degree of freedom in (a), and prove a new matrix inequality that gives us a better understanding of (b). Together, these techincal advances lead to a width-independent, parallel, simpler, and faster solver for positive SDPs.

## 1.1 Roadmap

We introduce our new matrix inequality and discuss about its connection to positive SDP in Section 2. Next in Section 3 we describe our algorithm `PosSDPSolver`. In Section 4, we define an objective $f_\mu(x)$ and relates it to positive SDP. In Section 5 and Section 6 respectively, we describe the convergence analyses for the packing and the covering SDPs.

## 2 Some False and Some True Inequalities in Matrix Algebra

We denote by $A \bullet B = \text{Tr}(AB) = \text{Tr}(BA)$ the matrix inner product, and by $\|A\|_{\text{spe}}$ the spectral norm of a matrix $A$. If $X$ is symmetric, we use $e^X$ to denote its matrix exponential. We write $A \succeq 0$ if $A$ is positive semidefinite (PSD), and $A \succeq B$ if $A - B \succeq 0$.

**Some False Matrix Inequalities.** The following is the SDP version of a fundamental inequality that the positive LP solver of [4] relies on: for every symmetric matrix $\Psi$ and every $i \in [n]$,

$$A_i \bullet e^{\Psi+B} = (1 \pm O(\varepsilon)) \cdot A_i \bullet e^{\Psi} \text{ if } -\varepsilon I \preceq B \preceq \varepsilon I \ . \tag{2.1}$$

Unfortunately, this inequality is *false* in the general SDP case. It is straightforward to check that it holds when all matrices involved are diagonal.

Similarly, here is another SDP inequality, whose LP version is crucial to to many positive LP solvers [8–10, 36, 37]. It is the following monotonocity statement: for every symmetric matrix $\Psi$ and every $i \in [n]$,

$$A_i \bullet e^{\Psi+B} \geq A_i \bullet e^{\Psi} \text{ if } B \succeq 0 \ .$$

However, this inequality is again *false.*

Unfortunately, these false matrix facts have found their ways in the positive SDP solvers proposed in [19, 30]. It is not clear at this point if these analyses can be fixed [29, 35].[3] Both the inequalities above become true if $\Psi$ and $B$ commute. This is precisely why the aforementioned positive LP solvers are correct.

**Our New Approach.** In this section, we shall prove that

$$B \bullet e^{\Psi+B} = (1 \pm O(\varepsilon)) \cdot B \bullet e^{\Psi} \text{ as long as } \varepsilon I \succeq B \succeq 0 \text{ or } -\varepsilon I \preceq B \preceq 0. \tag{2.2}$$

This non-trivial matrix inequality holds *even if* $B$ and $\Psi$ are not commutable, and shall become important for our later proofs in Section 5.1. We shall prove this by first establishing an interesting extended form of the Lieb-Thirring inequality.

In 1976, Lieb and Thirring [22] proved that for every $A, B \succeq 0$ and every $r \geq 1$, it holds that $\text{Tr}(B^{1/2}A^{1/2}B^{1/2})^r \leq \text{Tr}(B^{r/2}A^{r/2}B^{r/2})$. This inequality is known as the Lieb-Thirring inequality and is famous for its applications in quantum mechanics and differential equations. Very recently, Allen-Zhu, Liao, and Orecchia have connected it to the online matrix optimization problems [2].

---

[3]The ArXiv version [31] of the paper of Peng and Tangwongsan [30] acknowledges the error. The error in the analysis of [19] lies in the proof of Lemma 8, where they use the fact that "$\text{local}_j(x)$ only increases". This is an instantiation of the second false inequality above.

In the special case of $r = 2$, the Lieb-Thirring inequality says that $\text{Tr}(B^{1/2}A^{1/2}B^{1/2})^2 \leq \text{Tr}(BAB)$. In this paper, we establish the following generalization of the Lieb-Thirring inequality, which turns out to be crucial for the convergence analysis of our positive SDP solver. To the best of our knowledge, this inequality has not appeared in the literature.

**Lemma 2.1** (Extended Lieb-Thirring Inequality). *Given $A \succ 0$, $B \succeq 0$ and $\alpha \in [0, 1]$, we have*

$$B^{1/2}A^{\alpha}B^{1/2} \bullet B^{1/2}A^{1-\alpha}B^{1/2} \leq \text{Tr}(BAB) .$$

Unlike the original proof of Lieb-Thirring inequality which relies on Epstein's concavity theorem, our proof of Lemma 2.1 relies on Lieb's concavity theorem:

**Proposition 2.2** (Lieb's concavity theorem). *For all $m \times n$ matrices $K$, and all $q, r$ such that $0 \leq q \leq 1$ and $0 \leq r \leq 1$, with $q + r \leq 1$, the function $F(A, B) \stackrel{\text{def}}{=} \text{Tr}(K^T A^q K B^r)$ is jointly concave over $(A, B)$, where $A$ (resp. $B$) is over the set of all $m \times m$ (resp. $n \times n$) positive definite matrices.*

*Proof of Lemma 2.1.* The inequality is obvious when $\alpha = 0$ or $\alpha = 1$, and therefore we shall assume without loss of generality that $\alpha \in (0, 1)$. In addition, we can assume without loss of generality that $B$ is diagonal: otherwise, one can apply an orthogonal transformation to make $B$ diagonal.

Let us write $A = A^D + A^0$, where $A^D$ is the diagonal part of $A$, and $A^0$ is the off-diagonal part of $A$. Define $A_\lambda \stackrel{\text{def}}{=} A^D + \lambda A^0 = \lambda A + (1 - \lambda)A^D$. It is clear from this definition that $A_\lambda \succeq 0$ for all $\lambda \in [0, 1]$. In fact, we notice that $A \succ 0$ implies $A^D$ is positive in all of its diagonal entries. As a consequence, there exists some constant $\varepsilon > 0$ such that $A_\lambda \succ 0$ even for all $\lambda \in [-\varepsilon, 1]$.

Now, consider two matrix-to-real functions $g(A) \stackrel{\text{def}}{=} B^{1/2}A^{\alpha}B^{1/2} \bullet B^{1/2}A^{1-\alpha}B^{1/2}$ and $h(A) \stackrel{\text{def}}{=} \text{Tr}(BAB)$. Since $g(A) = \text{Tr}(BA^{\alpha}BA^{1-\alpha})$, Lieb's concavity theorem (cf. Proposition 2.2) implies that $g(A)$ is concave in $A$ (over the positive definite cone). In contrast, $h(A)$ is simply a function that is linear in $A$. Therefore, $R(\lambda) \stackrel{\text{def}}{=} g(A_\lambda) - h(A_\lambda)$ is defined and concave over $\lambda \in [-\varepsilon, 1]$, and Lemma 2.1 is equivalent to saying that $R(1) \leq 0$.

We begin analyzing $R(\lambda)$ by noticing that $R(0) = g(A_0) - h(A_0) = 0$: this is a simple consequence of the fact that $B$, being a diagonal matrix, commutes with $A_0 = A^D$. Therefore, combined with the concavity of $R(\lambda)$, to prove $R(1) \leq 0$ it suffices to prove that $R(\lambda)$ is differentiable at $\lambda = 0$ and $R'(0) = 0$.

First of all, $M_1(\lambda) \stackrel{\text{def}}{=} (A_\lambda)^{\alpha}$ is differentiable at $\lambda = 0$ and its derivative at $\lambda = 0$ has zero diagonal entries. Indeed, using the representation $M_1(\lambda) = \frac{1}{\pi \csc(\alpha \pi)} \cdot \int_0^\infty x^{\alpha-1} \cdot A_\lambda (A_\lambda + xI)^{-1} dx$, one can verify that,

$$\frac{dM_1(\lambda)}{d\lambda}\Big|_{\lambda=0} = \frac{1}{\pi \csc(\alpha \pi)} \cdot \int_0^\infty x^{\alpha-1} \cdot \Big(\frac{dA_\lambda}{d\lambda}(A_\lambda + xI)^{-1} - A_\lambda(A_\lambda + xI)^{-1}\frac{dA_\lambda}{d\lambda}(A_\lambda + xI)^{-1}\Big)\Big|_{\lambda=0} dx$$

$$= \frac{1}{\pi \csc(\alpha \pi)} \cdot \int_0^\infty x^{\alpha-1} \cdot \Big(A^0(A^D + xI)^{-1} - A^D(A^D + xI)^{-1}A^0(A^D + xI)^{-1}\Big) dx .$$

Noticing in the above equality $A^0$ is a matrix with zero diagonal entries, while $(A^D + xI)^{-1}$ and $A^D(A^D + xI)^{-1}$ are both diagonal matrices. Therefore, $M_1'(0)$ is a matrix with zero diagonal entries.

Similarly, defining $M_2(\lambda) \stackrel{\text{def}}{=} (A_\lambda)^{1-\alpha}$ we have that $M_2(\lambda)$ is differentiable at $\lambda = 0$ and $M_2'(0)$ is a matrix with zero diagonal entries.

**Algorithm 1** `PosSDPSolver(A, ε)`

**Input:** $\mathbf{A} = (A_1, \ldots, A_n)$ where each $A_i \in \mathbb{R}^{m \times m}$ is PSD, and $\varepsilon \in (0, 1/10]$.
**Output:** nonnegative vector $x \in \mathbb{R}^n_{\geq 0}$ and PSD matrix $Y \in \mathbb{R}^{m \times m}$.
1: $\mu \leftarrow \frac{\varepsilon}{4 \log(nm/\varepsilon)}$ and $\alpha \leftarrow \frac{\varepsilon \mu}{4}$. ⋄ *parameters*
2: $x_i^{(0)} \leftarrow \frac{1 - \varepsilon/2}{n \|A_i\|_{\mathsf{spe}}}$ for all $i \in [n]$. ⋄ *initial vector* $x^{(0)}$
3: $T \leftarrow \frac{8 \log(2n)}{\alpha \varepsilon}$. ⋄ *number of iterations*
4: **for** $k \leftarrow 0$ **to** $T - 1$ **do**
5:    Randomly choose $\mathbb{T}^{(k)}$ to be either $\mathbb{T}_-$ or $\mathbb{T}_+$, each with probability half.
6:    **for** $i \leftarrow 1$ **to** $n$ **do**
7:      Compute the feedback $v_i \leftarrow e^{\frac{1}{\mu}(\sum_{i \in [n]} x_i A_i - I)} \bullet A_i - 1$
8:      Perform an update: $x_i^{(k+1)} \leftarrow x_i^{(k)} \cdot e^{-\alpha \cdot \mathbb{T}^{(k)}(v_i)}$.
9:    **end for**
10: **end for**
11: **return** $\frac{x^{(T)}}{1+\varepsilon}$ and $\frac{\overline{Y}}{1-2\varepsilon}$, where $\overline{Y} \stackrel{\text{def}}{=} \sum_{i=0}^{T-1} Y(x^{(k)})$. ⋄ *recall that* $Y(x) \stackrel{\text{def}}{=} e^{\frac{1}{\mu}(\sum_{i \in [n]} x_i A_i - I)}$

Finally, we can compute that

$$R'(0) = \frac{d\big(B^{1/2}(A_\lambda)^\alpha B^{1/2} \bullet B^{1/2}(A_\lambda)^{1-\alpha} B^{1/2}\big)}{d\lambda}\bigg|_{\lambda=0} - \frac{d(B^2 \bullet A_\lambda)}{d\lambda}\bigg|_{\lambda=0}$$
$$= B^{1/2} M_1'(0) B^{1/2} \bullet B^{1/2}(A^D)^{1-\alpha} B^{1/2} + B^{1/2}(A^D)^\alpha B^{1/2} \bullet B^{1/2} M_2'(0) B^{1/2} - B^2 \bullet A^0 \ .$$

Clearly, this means $R'(0) = 0$ because $M_1'(0)$, $M_2'(0)$ and $A^0$ are all matrices with zero diagonal entries, and $B$ and $A^D$ are diagonal matrices. □

Our extended Lieb-Thirring inequality immediately yields the following monotonicity property on matrix exponential, which is a formal statement of (2.2). Its proof is deferred to Appendix A.

**Lemma 2.3.** *Given PSD matrix $A$ satisfying $\varepsilon I \succeq A \succeq 0$ and symmetric matrix $\Psi$, define function $f(t) \stackrel{\text{def}}{=} A \bullet e^{\Psi + tA}$ over real values $t$. Then, $0 \leq f'(t) \leq \varepsilon A \bullet e^{\Psi + tA} = \varepsilon f(t)$ for all $t$. As a result:*
*(a) $f(t) \leq f(0) \cdot e^{\varepsilon t}$ for all $t \geq 0$, and*
*(b) $f(t) \geq f(0) \cdot e^{\varepsilon t}$ for all $t \leq 0$.*

## 3 Our Algorithm

Our algorithm `PosSDPSolver(A, ε)` runs only in $T = O(\frac{\log n \cdot \log(nm/\varepsilon)}{\varepsilon^3})$ parallelizable iterations. We iteratively update $x$ so as to maximize $\mathbb{1}^T x$, while keeping the approximate feasibility $\sum_i x_i A_i \preceq (1+\varepsilon) I$. At each iteration $k$, we compute a feedback vector $v$ so that $v_i = e^{\frac{1}{\mu}(\sum_{i \in [n]} x_i A_i - I)} \bullet A_i - 1 \in [-1, \infty)$, and perform a multiplicative update $x_i \leftarrow x_i \cdot e^{-\alpha \cdot \mathbb{T}(v_i)}$. Here, $\mathbb{T}(\cdot)$ is randomly chosen (for each iteration $k$) as either $\mathbb{T}_-$ or $\mathbb{T}_+$, defined as follows:

**Definition 3.1.** *The thresholding functions $\mathbb{T}_-, \mathbb{T}_+ \colon [-1, \infty) \to [-1, 1]$ are defined as follows*

$$\mathbb{T}_-(v) \stackrel{\text{def}}{=} \begin{cases} 0, & v \in [-\varepsilon, \infty); \\ v, & v \in [-1, -\varepsilon). \end{cases} \quad \text{and} \quad \mathbb{T}_+(v) \stackrel{\text{def}}{=} \begin{cases} 0, & v \in [-1, \varepsilon]; \\ v, & v \in (\varepsilon, 1]; \\ 1, & v > 1. \end{cases}$$

Note that if $\mathbb{T} = \mathbb{T}_-$ then the variables of $x$ monotonically non-decreases, and vice versa.

**Remark 3.2** (Matrix Exponentials). Matrix exponential computations are required by all width-independent positive SDP solvers, and dominate the complexity of each algorithmic iteration. Like in previous solvers, it is a simple exercise to verify that our entire analysis in this paper continues to hold, though with a worsen constant, if we are only computing the values $v_i = e^{\frac{1}{\mu}(\sum_{i\in[n]} x_i A_i - I)} \bullet A_i$ up to a $1 \pm \varepsilon/2$ multiplicative factor. Therefore, for simplicity's sake, in this paper we assume that the matrix exponentials can be computed exactly. Note that the $1 \pm \varepsilon/2$ approximate computations of $e^{\frac{1}{\mu}(\sum_{i\in[n]} x_i A_i - I)} \bullet A_i$ for *all* $i \in [n]$ can be performed in polylog parallel iterations.[4]

We summarize our theorem as follows.

> **Theorem 3.4** (Positive SDP). *Letting $(x, Y) = \texttt{PosSDPSolver}(\mathbf{A}, \varepsilon)$, we have that with at least a constant probability*
> - *$x$ is a $(1 - O(\varepsilon))$-approximate solution for the packing SDP (1.1),*
> - *$Y$ is a $(1 + O(\varepsilon))$-approximate solution for the covering SDP (1.2), and*
> - *the number of iterations for $\texttt{PosSDPSolver}$ is $T = O(\log n \cdot \log(nm/\varepsilon) \cdot \varepsilon^{-3})$.*
>
> *If each $A_i = Q_i Q_i^T$ is preprocessed into its Cholesky decomposition, each iteration can be implemented in $O(\log^2(nm)/\varepsilon)$ parallel depth.*

## 4 The Convex Objective

We define the following convex objective for the positive SDP problem. It is completely analogous to its LP variant introduced in [4], and therefore we state its properties without proof.

**Definition 4.1.** *Letting parameter $\mu \stackrel{\text{def}}{=} \frac{\varepsilon}{4 \log(nm/\varepsilon)}$, we define the smoothed objective $f_\mu(x)$ as*

$$f_\mu(x) \stackrel{\text{def}}{=} \mu \cdot \mathrm{Tr}\big(e^{\frac{1}{\mu}(\sum_{i\in[n]} x_i A_i - I)}\big) - \mathbb{1}^T x \ .$$

We want to study the minimization problem on $f_\mu(x)$ over all $x \geq 0$. This objective $f_\mu(x)$ captures the packing SDP because, on one hand we want to minimize $-\mathbb{1}^T x$ so as to maximize $\mathbb{1}^T x$, and on the other hand the exponential penalty function says if $\sum_{i\in[n]} x_i A_i \preceq (1+\varepsilon)I$ is violated, a large positive penalty is introduced.

**Proposition 4.2.**
  (a) $\mathsf{OPT} \in [1, n]$.
  (b) *Letting $x = (1 - \varepsilon/2)x^* \geq 0$, we have $f_\mu(x) \leq -(1 - \varepsilon)\mathsf{OPT}$.*
  (c) *Letting $x^{(0)} \geq 0$ be such that $x_i^{(0)} = \frac{1 - \varepsilon/2}{n\|A_i\|_{\mathsf{spe}}}$ for each $i \in [n]$, we have $f_\mu(x^{(0)}) \leq -\frac{1-\varepsilon}{n}$.*
  (d) *For any $x \geq 0$ satisfying $f_\mu(x) \leq 0$, we have $\sum_{i\in[n]} x_i A_i \preceq (1+\varepsilon)I$ and thus $\mathbb{1}^T x \leq (1+\varepsilon)\mathsf{OPT}$.*

---

[4]More precisely, when each $A_i = Q_i Q_i^T$ is presented in its Cholesky decomposition, we have

**Theorem 3.3** ([30]). *Given an $m \times m$ PSD matrix $\Phi$ with $p$ non-zero entries and $\|\Phi\|_{\mathsf{spe}} \leq \kappa$, and given $m \times m$ matrices $\{A_1, \ldots, A_n\}$ in the form of $A_i = Q_i Q_i^T$ where the total non-zero entries across all $Q_i$ is $q$. Then, there exists an algorithm that computes $e^\Phi \bullet A_i$ for all $i \in [n]$ up to a $(1 \pm \varepsilon)$ factor in*

$$O\Big(\max\big\{\kappa, \log \frac{1}{\varepsilon}\big\} \log m + \log \log m\Big) \ \text{depth} \quad \text{and} \quad O\Big(\frac{1}{\varepsilon^2}\Big(\max\big\{\kappa, \log \frac{1}{\varepsilon}\big\} \cdot p + q\Big) \log m\Big) \ \text{work}$$

Since one can verify that $\|\Phi\|_{\mathsf{spe}} \leq \kappa \stackrel{\text{def}}{=} 1/\mu = O(\log(nm/\varepsilon)/\varepsilon)$ in our case, each iteration of $\texttt{PosSDPSolver}$ can be implemented to run in $O(\log^2(nm)/\varepsilon)$ parallel time. (Here, we can safely assume that $\varepsilon > 1/(nm)^{O(1)}$; if $\varepsilon$ is smaller than $1/(nm)^{O(1)}$, one should use for instance Interior Point Method to solve the given SDP instead.)

(e) If $x \geq 0$ satisfies $f_\mu(x) \leq -(1-O(\varepsilon))\mathsf{OPT}$, then $\frac{1}{1+\varepsilon}x$ is a $(1-O(\varepsilon))$-approximate solution for the packing SDP.

(f) The gradient of $f_\mu(x)$ can be written as

$$\nabla f_\mu(x) = (A_1 \bullet Y(x), \ldots, A_n \bullet Y(x)) - \mathbb{1} \quad \text{where} \quad Y(x) \stackrel{\text{def}}{=} e^{\frac{1}{\mu}(\sum_{i \in [n]} x_i A_i - I)} \qquad (4.1)$$

## 5 Convergence Analysis for Packing SDP

Throughout this paper, we use superscript $x^{(k)}$ to represent vector $x$ at iteration $k$, and subscript $x_i$ to represent the $i$-th coordinate of vector $x$. Our convergence analysis is divided into three steps, and the first step is the main technical difference between this paper and its LP variant [4].

**Step I: Gradient Descent.** We interpret (see Section 5.1 for details) each update $x_i^{(k+1)} \leftarrow x_i^{(k)} \cdot e^{-\alpha \cdot \mathbb{T}^{(k)}(v_i)}$ as a gradient descent step,[5] and show that the objective $f_\mu(x)$ monotonically decreases between consecutive iterations:

> **Lemma 5.1** (Gradient Descent). *For every iteration $k = 0, \ldots, T-1$ in* `PosSDPSolver`, *the objective $f_\mu(x)$ does not increases: $f_\mu(x^{(k)}) - f_\mu(x^{(k+1)}) \geq 0$. Combining this with Proposition 4.2.c, we have $f_\mu(x^{(k)}) \leq 0$ for all $k$.*
>
> *In addition, letting $B^{(k)} \subseteq [n]$ be the set of indices $i$ such that $\nabla_i f_\mu(x^{(k)}) \geq 1$, then*
>
> $$f_\mu(x^{(k)}) - \mathbf{E}[f_\mu(x^{(k+1)})] \geq \frac{\alpha}{4} \cdot \sum_{i \in B^{(k)}} x_i^{(k)} \cdot \nabla_i f_\mu(x^{(k)}) \geq 0 \ .$$
>
> *Above, the expectation is over the random choice of $\mathbb{T}^{(k)}$ at iteration $k$.*

We remark here that Lemma 5.1 does *not* follow from any classical theory of gradient descent because our objective $f_\mu(x)$ is simply not smooth in the positive orthant. Neither does Lemma 5.1 follow from the so-called "multiplicative Lipschitz gradient property" introduced in [4], because the fundamental property that the work [4] replies on, "$\nabla_i f_\mu(x)$ increases as $x$ decreases, and vice versa", no longer holds in the SDP case. This is also one of the major reasons that the results of [19, 30] fail to produce any theoretical guarantee.

Our proof of Lemma 5.1 crucially relies on two key properties. First, the sign-consistent and random choice of $\mathbb{T}^{(k)}$ ensures that $x$ either only increases or only decreases at a single iteration $k$. Second, our new matrix inequality introduced in Section 2 ensures that "$\nabla_i f_\mu(x)$ increases *in an average sense* as $x$ decreases". We defer the technical proof of Lemma 5.1 to Section 5.1.

**Step II: Mirror Descent.** It is not hard to show, and in fact proven in [4] for a slightly different variant, that each update $x_i^{(k+1)} \leftarrow x_i^{(k)} \cdot e^{-\alpha \cdot \mathbb{T}^{(k)}(v_i)}$ can also be viewed as a mirror-descent step.

A *mirror descent step* in optimization is any step from $x$ to $x'$ that is of the form $x' \leftarrow \arg\min_z \{V_x(z) + \langle \alpha \nabla f(x), z - x \rangle\}$. Here, $\alpha > 0$ is some step length, and $V_x(\widetilde{x}) = w(\widetilde{x}) - \langle \nabla w(x), \widetilde{x} - x \rangle - w(x)$ is the Bregman divergence of some convex *distance generating function* $w(x)$. In this paper, we pick $w(x) \stackrel{\text{def}}{=} \sum_{i \in [n]} x_i \log x_i - x_i$ to be the generalized entropy function, and accordingly,

$$\text{for every } x, \widetilde{x} \geq 0, \quad V_x(\widetilde{x}) \stackrel{\text{def}}{=} \sum_{i \in [n]} \big(\widetilde{x}_i \log \tfrac{\widetilde{x}_i}{x_i} + x_i - \widetilde{x}_i\big) \ .$$

The next lemma easily follows from the general theory of mirror descent. Since its proof has essentially appeared in [4, Lemma 3.3], we prove it in Section B.3 only for the sake of completeness.

---

[5]To be clear, in some literature, the gradient descent is referred only to $x \leftarrow x - c \cdot \nabla f(x)$ for some constant $c$. In this paper, we adopt the more general notion, and refer it to any step that directly decreases $f(x)$.

**Lemma 5.2** (Mirror Descent). *Letting $\gamma \in [-1, 1]^n$ be defined as $\gamma_i = \mathbb{T}(\nabla_i f_\mu(x^{(k)}))$, we have that for any $u \geq 0$,*
$$\langle \alpha \gamma, x^{(k)} - u \rangle \leq \alpha^2 \mathsf{OPT} + V_{x^{(k)}}(u) - V_{x^{(k+1)}}(u) \ .$$

**Step III: Coupling.** Finally, as formally argued in Section B.2, the two lemmas above can be naturally combined, yielding the following bound:

**Lemma 5.3** (Coupling). *For any $u \geq 0$ and $k = 0, \ldots, T - 1$, we have*
$$\alpha(f_\mu(x^{(k)}) - f_\mu(u)) \leq \langle \alpha \nabla f_\mu(x^{(k)}), x^{(k)} - u \rangle$$
$$\leq 4(f_\mu(x^{(k)}) - \mathbf{E}[f_\mu(x^{(k+1)})]) + 2(V_{x^{(k)}}(u) - \mathbf{E}[V_{x^{(k+1)}}(u)]) + \alpha \cdot 2\varepsilon \mathsf{OPT} + \alpha \cdot \varepsilon \mathbb{1}^T u \ .$$

*Above, the expectation is over the random choice of $\mathbb{T}^{(k)}$ at iteration $k$.*

The proof of Lemma 5.3 relies on a decomposition of the gradient $\nabla_i f_\mu(x^{(k)})$ into four components $\nabla_i f_\mu(x^{(k)}) = \xi_i^+ + \xi_i^- + \eta_i + \zeta_i$, where $\xi_i^+ \in [0, 1]$, $\xi_i^- \in [-1, 0]$, $\eta_i \in [0, \infty)$, and $\zeta_i \in [-\varepsilon, \varepsilon]$. This is a main difference that distinguishes our proof from [4]: we need to decompose the $\xi_i$ part into a positive and a negative terms, and then apply Lemma 5.2 twice.

**Putting All Together.** By telescoping the inequality in Lemma 5.3, one can obtain the following final theorem for packing SDP. Its proof is only slightly different from that of [4, Theorem 3.5] due to the special treatment of the randomness, and deferred to Section B.4.

**Theorem 5.4** (Packing SDP). *For $T \geq \frac{8 \log(2n)}{\alpha \varepsilon} = \Omega(\frac{\log n \cdot \log(nm/\varepsilon)}{\varepsilon^3})$, we have that $\mathbf{E}[f_\mu(x^{(T)})] \leq -(1 - 5\varepsilon)\mathsf{OPT}$. As a consequence, $\mathtt{PosSDPSolver}(A, \varepsilon)$ produces an output $x = \frac{x^{(T)}}{1 + \varepsilon}$ that is a $(1 - O(\varepsilon))$-approximate solution for the packing SDP (1.1) with at least a constant probability.*

## 5.1 The Gradient Descent Lemma

In this subsection we view our update $x^{(k)} \to x^{(k+1)}$ as a gradient-descent step and prove Lemma 5.1. We begin by observing that each $x_i$ is changed by a factor of at most $1 \pm 4\alpha/3$ per iteration:

**Fact 5.5.** *We always have $x_i^{(k+1)} \in x_i^{(k)} \cdot [1 - 4\alpha/3, 1 + 4\alpha/3]$.*

*Proof.* We can always write $x_i^{(k+1)} = x_i^{(k)} \cdot e^t$ for some $t \in [-\alpha, \alpha] \subseteq [-1/4, 1/4]$. According to the fact that $e^t \leq 1 + 4t/3$ for $t \in [0, 1/4]$ and $e^t \geq 1 - t \geq 1 - 4t/3$ for $t \in [-1/4, 0]$, we must have $x_i^{(k+1)} \in x_i^{(k)} \cdot [1 - 4\alpha/3, 1 + 4\alpha/3]$. □

*Proof of Lemma 5.1.* We prove by induction. Suppose that Lemma 5.1 is true for all indices less than $k$. This implies, in particular, that $f_\mu(x^{(k)}) \leq f_\mu(x^{(k-1)}) \leq \cdots \leq f_\mu(x^{(0)}) \leq 0$.

There are two cases to consider at iteration $k$: (1) if we choose $\mathbb{T}_-(\cdot)$ and (2) if we choose $\mathbb{T}_+(\cdot)$. Each of them happens with probability $1/2$.

In the first case, that is, if we choose $\mathbb{T}_-(\cdot)$, we have the property that our vector does not decrease: that is, $x_i^{(k+1)} \geq x_i^{(k)}$ for every $i \in [n]$. We compute the objective difference by the standard integral over gradients:

$$f_\mu(x^{(k)}) - f_\mu(x^{(k+1)}) = \int_0^1 \left\langle \nabla f_\mu\big(x^{(k)} + \tau(x^{(k+1)} - x^{(k)})\big), x^{(k)} - x^{(k+1)} \right\rangle d\tau$$

$$= \mathbb{1}^T x^{(k+1)} - \mathbb{1}^T x^{(k)} + \int_0^1 \left( e^{\frac{1}{\mu}(\sum_{i\in[n]} x_i^{(k)} A_i - I + \tau \sum_{i\in[n]}(x_i^{(k+1)} - x_i^{(k)}) A_i)} \bullet \sum_i (x_i^{(k)} - x_i^{(k+1)}) A_i \right) d\tau$$

$$= \mathbb{1}^T x^{(k+1)} - \mathbb{1}^T x^{(k)} - \mu \int_0^1 B \bullet e^{\Psi + \tau B} d\tau \;, \tag{5.1}$$

where in the last equality we have defined $\Psi \stackrel{\text{def}}{=} \frac{1}{\mu}(\sum_{i\in[n]} x_i^{(k)} A_i - I)$ and $B \stackrel{\text{def}}{=} \frac{1}{\mu} \sum_{i\in[n]}(x_i^{(k+1)} - x_i^{(k)}) A_i \succeq 0$.

Notice that $f_\mu(x^{(k)}) \leq 0$ together with Proposition 4.2.d tells us that $\sum_{i\in[n]} x_i^{(k)} A_i \preceq (1+\varepsilon) I$. Combining it with Fact 5.5 we have $\sum_{i\in[n]} (x_i^{(k+1)} - x_i^{(k)}) A_i \preceq \frac{4\alpha}{3}(1+\varepsilon) I \preceq \frac{5\alpha}{3} I$ and therefore $B \preceq \frac{5\alpha}{3\mu} I = \frac{5\varepsilon}{12} I$. Applying Lemma 2.3.a with $B \preceq \frac{5\varepsilon}{12} I$ to (5.1), we have

$$f_\mu(x^{(k)}) - f_\mu(x^{(k+1)}) \geq \mathbb{1}^T x^{(k+1)} - \mathbb{1}^T x^{(k)} - \mu \int_0^1 B \bullet e^\Psi \cdot e^{5\varepsilon\tau/12} d\tau$$

$$\geq \mathbb{1}^T x^{(k+1)} - \mathbb{1}^T x^{(k)} - (1+\varepsilon/4) \mu B \bullet e^\Psi \;.$$

Recall that, for each $i \in [n]$ satisfying $x_i^{(k+1)} \neq x_i^{(k)}$, we must have $e^\Psi \bullet A_i - 1 < -\varepsilon$ by the definition of $\mathbb{T}_-(\cdot)$. Therefore, multiplying both sides by $x_i^{(k+1)} - x_i^{(k)} \geq 0$ and summing up over $i \in [n]$, we obtain

$$\mu B \bullet e^\Psi = e^\Psi \bullet \Big(\sum_{i\in[n]} (x_i^{(k+1)} - x_i^{(k)}) A_i\Big) \leq (1-\varepsilon)(\mathbb{1}^T x^{(k+1)} - \mathbb{1}^T x^{(k)}) \;.$$

This further implies that (after some careful term rearranging)

$$\mathbb{1}^T x^{(k+1)} - \mathbb{1}^T x^{(k)} - (1+\varepsilon/4) \mu B \bullet e^\Psi \geq \tfrac{3}{4}(\mathbb{1}^T x^{(k+1)} - \mathbb{1}^T x^{(k)} - \mu B \bullet e^\Psi)$$
$$= \tfrac{3}{4} \langle \nabla f_\mu(x^{(k)}), x^{(k)} - x^{(k+1)} \rangle \geq 0 \;.$$

Above, the last inequality is again by our definition of $\mathbb{T}_-$: for each $i \in [n]$ satisfying $x_i^{(k)} \neq x_i^{(k+1)}$, it must satisfy that $\nabla_i f_\mu(x^{(k)}) < -\varepsilon$ and $x_i^{(k)} \leq x_i^{(k+1)}$. In conclusion, we arrive at the inequality

$$f_\mu(x^{(k)}) - f_\mu(x^{(k+1)}) \geq \frac{3}{4} \langle \nabla f_\mu(x^{(k)}), x^{(k)} - x^{(k+1)} \rangle \geq 0 \;.$$

In the case when $\mathbb{T}_+$ is chosen, a symmetric argument (although replacing the use of Lemma 2.3.a with Lemma 2.3.b and using slightly different constants, see Appendix B.1) yields that

$$f_\mu(x^{(k)}) - f_\mu(x^{(k+1)}) \geq \frac{2}{3} \langle \nabla f_\mu(x^{(k)}), x^{(k)} - x^{(k+1)} \rangle$$
$$\geq \frac{2}{3} \sum_{i \in B^{(k)}} \nabla_i f_\mu(x^{(k)}) \cdot (x_i^{(k)} - x_i^{(k+1)}) \;.$$

Above, the second inequality is because for each $i \in [n]$ satisfying $x_i^{(k)} \neq x_i^{(k+1)}$, it must satisfy that $\nabla_i f_\mu(x^{(k)}) > \varepsilon$ and $x_i^{(k)} \geq x_i^{(k+1)}$. Next, observe that for each coordinate $i \in B^{(k)}$ we have $x_i^{(k+1)} = x_i^{(k)} \cdot e^{-\alpha} \leq (1 - 0.9\alpha) x_i^{(k)}$ for our choice of $\alpha$. Plugging this into the inequality above, we arrive at the inequality

$$f_\mu(x^{(k)}) - f_\mu(x^{(k+1)}) \geq \frac{2}{3} \cdot 0.9\alpha \sum_{i \in B^{(k)}} \nabla_i f_\mu(x^{(k)}) \cdot x_i^{(k)} \geq \frac{\alpha}{2} \sum_{i \in B^{(k)}} \nabla_i f_\mu(x^{(k)}) \cdot x_i^{(k)} \geq 0 \enspace .$$

Finally, combining the two cases above, we conclude that

$$f_\mu(x^{(k)}) - \mathbf{E}[f_\mu(x^{(k+1)})] \geq \frac{\alpha}{4} \sum_{i \in B^{(k)}} \nabla_i f_\mu(x^{(k)}) \cdot x_i^{(k)} \enspace . \qquad \square$$

## 6 Convergence Analysis for Covering SDP

We have seen in Section 5 that a vector $x \geq 0$ satisfying $f_\mu(x) \approx -\mathsf{OPT}$ yields an approximate solution to the packing SDP (1.1). However, this vector $x$ itself gives no information about the solution to the covering SDP (1.2).

In this section, we show that, defining $\overline{Y} \stackrel{\text{def}}{=} \sum_{i=0}^{T-1} Y(x^{(k)})$ where $Y(x) \stackrel{\text{def}}{=} e^{\frac{1}{\mu}(\sum_{i \in [n]} x_i A_i - I)}$, then $\frac{\overline{Y}}{1-2\varepsilon}$ is a $(1 + O(\varepsilon))$-approximate solution to the covering SDP (1.2) with at least a constant probability. Therefore, `PosSDPSolver(A, ε)` is an algorithm that simultaneously solves both the primal and the dual side of the positive SDP problem.

Our proof can be divided into two parts. First, using similar proof techniques as in [4], one can show that $\overline{Y}$ satisfies the *approximate optimality*, at least in an expected sense. We prove this lemma below in Appendix C only for the sake of completeness.

**Lemma 6.1.** *For any $T \geq \frac{8}{\alpha\varepsilon} = \Omega(\frac{\log(nm/\varepsilon)}{\varepsilon^3})$, we have that $\mathbf{E}[\mathrm{Tr}(\overline{Y})] \leq (1 + 7\varepsilon)\mathsf{OPT}$.*

In the second part, we wish to show that $\overline{Y}$ satisfies the *approximate feasibility* as well, that is, $A_i \bullet \overline{Y} \leq 1 + O(\varepsilon)$ for all $i \in [n]$. However, we encounter two difficulties:
- First, a similar analysis as in [4] would only imply that the expected matrix $\mathbf{E}[\overline{Y}]$ satisfies such approximate feasibility, rather than $\overline{Y}$. By Markov's inequality, this only suggests that for each (<u>rather than for all</u>) $i \in [n]$, $A_i \bullet \overline{Y} \leq 1 + O(\varepsilon)$ holds with constant probability.[6]
- Second, the analysis in [4] does not directly imply that $\overline{Y}$ is approximately feasible. Instead, one has to modify $\overline{Y}$ in a non-trivial manner which is very unpleasant in practice.

Due to the above difficulties, we propose in this paper a fundamentally different, yet much simpler analysis for proving the approximate feasibility. This is deferred to Appendix C.

**Lemma 6.2.** *For any $T \geq \frac{8}{\alpha\varepsilon}$, with probability at least $1 - \frac{\varepsilon}{100}$ we have $A_i \bullet \overline{Y} \geq 1 - 2\varepsilon$ for all $i \in [n]$.*

It is now easy to see that Lemma 6.1 and Lemma 6.2 together imply that

> **Corollary 6.3** (Covering SDP). *With at least a constant probability, we have*
> 
> $$\forall i \in [n], A_i \bullet \overline{Y} \geq 1 - 2\varepsilon \quad \text{and} \quad \mathrm{Tr}(\overline{Y}) \leq 1 + O(\varepsilon)\mathsf{OPT} \enspace .$$
> 
> *Therefore, $\frac{\overline{Y}}{1-2\varepsilon}$ gives a $(1 + O(\varepsilon))$-approximate solution to the covering SDP (1.2).*

---

[6]Previously, the first and third authors of this paper have tried to bypass this difficulty using a dual smoothed objective in the LP case [3]. However, their analysis is more involved and loses a factor of $\varepsilon^{0.5}$ in the running time.

## Acknowledgements

We thank Richard Peng for helpful conversations. This material is based upon work partly supported by the National Science Foundation under Grant CCF-1319460.

# Appendix

## A  Missing Proofs for Section 2

We need the following chain rule for the derivative of matrix exponential:

**Proposition A.1** ([34]). *If $X(t)$ is a differentiable function from reals to symmetric matrices,*

$$\frac{d}{dt}e^{X(t)} = \int_{\alpha=0}^{1} e^{\alpha X(t)} \frac{dX(t)}{dt} e^{(1-\alpha)X(t)} d\alpha \ .$$

*Proof of Lemma 2.3.* According to Proposition A.1, we have

$$f'(t) = A \bullet \int_{\alpha=0}^{1} e^{\alpha(\Psi+tA)} A e^{(1-\alpha)(\Psi+tA)} d\alpha$$

Suppose further that $A = PP^T$. Then, we can write

$$f'(t) = \int_{\alpha=0}^{1} \text{Tr}\Big(P^T e^{\alpha(\Psi+tA)} PP^T e^{(1-\alpha)(\Psi+tA)} P\Big) d\alpha$$

However, since $P^T e^{\alpha(\Psi+tA)} P \succeq 0$ and $P^T e^{(1-\alpha)(\Psi+tA)} P \succeq 0$, we conclude that $P^T e^{\alpha(\Psi+tA)} P \bullet P^T e^{(1-\alpha)(\Psi+tA)} P \geq 0$ and therefore $f'(t) \geq 0$ for all reals $t$.

Next, applying Lemma 2.1 we have that

$$f'(t) = \int_{\alpha=0}^{1} \text{Tr}\Big(A e^{\alpha(\Psi+tA)} A e^{(1-\alpha)(\Psi+tA)}\Big) d\alpha \leq \int_{\alpha=0}^{1} \text{Tr}\Big(A^2 e^{\Psi+tA}\Big) d\alpha = A^2 \bullet e^{\Psi+tA} \leq \varepsilon A \bullet e^{\Psi+tA} \ .$$

$\square$

## B  Missing Proofs for Section 5

### B.1  The Gradient Descent Lemma

In this section, we provide the detailed analysis of the symmetric case (i.e., when $\mathbb{T}_+$ is chosen) in the proof for Lemma 5.1.

Notice that $f_\mu(x^{(k)}) \leq 0$ together with Proposition 4.2.d tells us that $\sum_{i\in[n]} x_i^{(k)} A_i \preceq (1+\varepsilon)I$. Combining it with Fact 5.5 we have $\sum_{i\in[n]} \big(x_i^{(k+1)} - x_i^{(k)}\big) A_i \succeq -\frac{4\alpha}{3}(1+\varepsilon)I \succeq -\frac{5\alpha}{3}I$ and therefore $0 \succeq B \succeq -\frac{5\alpha}{3\mu}I = -\frac{5\varepsilon}{12}I$. Applying Lemma 2.3.b with $0 \succeq B \succeq -\frac{5\varepsilon}{12}I$ to (5.1), we have

$$f_\mu(x^{(k)}) - f_\mu(x^{(k+1)}) \geq \mathbb{1}^T x^{(k+1)} - \mathbb{1}^T x^{(k)} - \mu \int_0^1 B \bullet e^\Psi \cdot e^{-5\varepsilon\tau/12} d\tau$$

$$\geq \mathbb{1}^T x^{(k+1)} - \mathbb{1}^T x^{(k)} - (1-\varepsilon/4)\mu B \bullet e^\Psi \ .$$

Recall that, for each $i \in [n]$ satisfying $x_i^{(k+1)} \neq x_i^{(k)}$, we must have $e^\Psi \bullet A_i - 1 > \varepsilon$ by the definition of $\mathbb{T}_+(\cdot)$. Therefore, multiplying both sides by $x_i^{(k+1)} - x_i^{(k)} \leq 0$ and summing up over $i \in [n]$, we obtain

$$\mu B \bullet e^\Psi = e^\Psi \bullet \big(\sum_{i \in [n]} (x_i^{(k+1)} - x_i^{(k)}) A_i\big) \leq (1+\varepsilon)(\mathbb{1}^T x^{(k+1)} - \mathbb{1}^T x^{(k)}) \ .$$

This further implies that (after some careful term rearranging)[7]

$$\mathbb{1}^T x^{(k+1)} - \mathbb{1}^T x^{(k)} - (1-\varepsilon/4)\mu B \bullet e^\Psi \geq \frac{2}{3}(\mathbb{1}^T x^{(k+1)} - \mathbb{1}^T x^{(k)} - \mu B \bullet e^\Psi)$$
$$= \frac{2}{3}\langle \nabla f_\mu(x^{(k)}), x^{(k)} - x^{(k+1)}\rangle \geq 0 \ .$$

Above, the last inequality is again by our definition of $\mathbb{T}_-$: for each $i \in [n]$ satisfying $x_i^{(k)} \neq x_i^{(k+1)}$, it must satisfy that $\nabla_i f_\mu(x^{(k)}) < -\varepsilon$ and $x_i^{(k)} \leq x_i^{(k+1)}$. In conclusion, we arrive at the inequality

$$f_\mu(x^{(k)}) - f_\mu(x^{(k+1)}) \geq \frac{2}{3}\langle \nabla f_\mu(x^{(k)}), x^{(k)} - x^{(k+1)}\rangle \geq 0 \ .$$

### B.2 The Coupling Lemma

The main idea in our proof to Lemma 5.3 is to divide the gradient vector $\nabla f(x) \in [-1, \infty)^n$ into four components, the component containing large coordinates (i.e., bigger than 1), the component containing positive small coordinates (i.e., in $(\varepsilon, 1]$), the component containing negative small coordinates (i.e., in $[-1, -\varepsilon)$), and the component containing negligible coordinates (i.e., in $[-\varepsilon, \varepsilon]$). The large gradients are to be taken care by the gradient descent lemma, the small (positive and negative) gradients are to be taken care by the mirror descent lemma. Formally,

*Proof of Lemma 5.3.* By convexity, the distance $f_\mu(x^{(k)}) - f_\mu(u)$ for an arbitrary $u \geq 0$ is upper bounded as follows:

$$\alpha(f_\mu(x^{(k)}) - f_\mu(u)) \leq \langle \alpha \nabla f_\mu(x^{(k)}), x^{(k)} - u\rangle$$
$$= \langle \alpha \eta^{(k)}, x^{(k)} - u\rangle + \langle \alpha \xi^{(k-)}, x^{(k)} - u\rangle + \langle \alpha \xi^{(k+)}, x^{(k)} - u\rangle + \langle \alpha \zeta^{(k)}, x^{(k)} - u\rangle \ ,$$
(B.1)

where
- $\xi_i^{(k-)} \stackrel{\text{def}}{=} \mathbb{T}_-(\nabla_i f_\mu(x^{(k)})) \in [-1, -\varepsilon)$ is the *truncated gradient*, capturing small negative coordinates.
- $\xi_i^{(k+)} \stackrel{\text{def}}{=} \mathbb{T}_+(\nabla_i f_\mu(x^{(k)})) \in (\varepsilon, 1]$ is the *truncated gradient*, capturing small positive coordinates.
- $\eta_i^{(k)} \stackrel{\text{def}}{=} \left\{ \begin{array}{ll} \nabla_i f_\mu(x^{(k)}) - 1, & \text{if } \nabla_i f_\mu(x^{(k)}) \geq 1; \\ 0, & \text{otherwise.} \end{array} \right\} \in [0, \infty)$, capturing the large coordinates.
- $\zeta_i^{(k)} \stackrel{\text{def}}{=} \left\{ \begin{array}{ll} \nabla_i f_\mu(x^{(k)}), & \text{if } \nabla_i f_\mu(x^{(k)}) \in [-\varepsilon, \varepsilon]; \\ 0, & \text{otherwise.} \end{array} \right\} \in [-\varepsilon, \varepsilon]$, capturing the negligible coordinates.

---
[7]Indeed, $\mu B \bullet e^\Psi \leq (1+\varepsilon)(\mathbb{1}^T x^{(k+1)} - \mathbb{1}^T x^{(k)})$ implies that $(1 - 3\varepsilon/4) \cdot \mu B \bullet e^\Psi \leq \mathbb{1}^T x^{(k+1)} - \mathbb{1}^T x^{(k)}$ because both sides are nonpositive and $1 - 3\varepsilon/4 \geq \frac{1}{1+\varepsilon}$ for our choice of $\varepsilon$. Multiplying both sides by $1/3$, we have that $(1/3 - \varepsilon/4) \cdot \mu B \bullet e^\Psi \leq (1/3) \cdot (\mathbb{1}^T x^{(k+1)} - \mathbb{1}^T x^{(k)})$. This is now equivalent to $\mathbb{1}^T x^{(k+1)} - \mathbb{1}^T x^{(k)} - (1-\varepsilon/4)\mu B \bullet e^\Psi \geq \frac{2}{3}(\mathbb{1}^T x^{(k+1)} - \mathbb{1}^T x^{(k)} - \mu B \bullet e^\Psi)$.

We analyze the four components of (B.1) one by one.

The $\zeta$ component is small: if $f_\mu(u) \leq 0$, we have

$$\langle \alpha \zeta^{(k)}, x^{(k)} - u \rangle \leq \alpha\varepsilon \cdot (\mathbb{1}^T x^{(k)} + \mathbb{1}^T u) \leq \alpha\varepsilon \cdot (1+\varepsilon)\mathsf{OPT} + \alpha\varepsilon \cdot \mathbb{1}^T u \tag{B.2}$$

where the last inequality is because $f_\mu(x^{(k)}) \leq 0$ from Lemma 5.1.

The $\eta$ component can be upper bounded with the help from Lemma 5.1 as follows. Note that $\eta_i^{(k)} \neq 0$ only if $i \in B^{(k)}$ (where recall from Lemma 5.1 that $B^{(k)}$ is the set of indices whose $\nabla_i f_\mu(x^{(k)})$ is no less than 1). In particular, if $i \in B^{(k)}$ we have $\eta_i^{(k)} = \nabla_i f_\mu(x^{(k)}) - 1 < \nabla_i f_\mu(x^{(k)})$, and thus Lemma 5.1 gives

$$\frac{4(f_\mu(x^{(k)}) - \mathbf{E}[f_\mu(x^{(k+1)})])}{\alpha} \geq \sum_{i \in B^{(k)}} x_i^{(k)} \cdot \nabla_i f_\mu(x^{(k)}) \geq \langle \eta^{(k)}, x^{(k)} \rangle$$
$$\implies \langle \alpha\eta^{(k)}, x^{(k)} - u \rangle \leq \langle \alpha\eta^{(k)}, x^{(k)} \rangle \leq 4(f_\mu(x^{(k)}) - \mathbf{E}[f_\mu(x^{(k+1)})])$$

Finally, the $\xi$ components are upper bounded by Lemma 5.2 as follows. Letting $\gamma = \xi^{(k-)}$ if $\mathbb{T}^{(k)} = \mathbb{T}_-$, and $\gamma = \xi^{(k+)}$ if $\mathbb{T}^{(k)} = \mathbb{T}_+$, we have that

$$\langle \alpha\xi^{(k-)}, x^{(k)} - u \rangle + \langle \alpha\xi^{(k+)}, x^{(k)} - u \rangle = 2\mathbf{E}[\langle \alpha\gamma, x^{(k)} - u \rangle] \leq 2\alpha^2 \mathsf{OPT} + 2V_{x^{(k)}}(u) - 2\mathbf{E}[V_{x^{(k+1)}}(u)] \ ,$$

where the expectation is over the random choice of $\mathbb{T}$ at iteration $k$.

Together, we obtain

$$\alpha(f_\mu(x^{(k)}) - f_\mu(u)) \leq \langle \alpha\eta^{(k)}, x^{(k)} - u \rangle + \langle \alpha\xi^{(k-)} + \alpha\xi^{(k+)}, x^{(k)} - u \rangle + \langle \alpha\zeta^{(k)}, x^{(k)} - u \rangle$$
$$\leq 4(f_\mu(x^{(k)}) - \mathbf{E}[f_\mu(x^{(k+1)})]) + 2\alpha^2 \mathsf{OPT} + 2V_{x^{(k)}}(u) - 2\mathbf{E}[V_{x^{(k+1)}}(u)] + \alpha\varepsilon \cdot (1+\varepsilon)\mathsf{OPT} + \alpha\varepsilon \mathbb{1}^T u$$
$$\leq 4(f_\mu(x^{(k)}) - \mathbf{E}[f_\mu(x^{(k+1)})]) + 2(V_{x^{(k)}}(u) - 2\mathbf{E}[V_{x^{(k+1)}}(u)]) + \alpha \cdot 2\varepsilon\mathsf{OPT} + \alpha \cdot \varepsilon \mathbb{1}^T u \ . \qquad \square$$

### B.3 The Mirror Descent Lemma

In this subsection, we are going to view our step $x^{(k)} \to x^{(k+1)}$ as a mirror descent step, and prove Lemma 5.2. We emphasize that this subsection is included in this paper only for the sake of completeness: it is almost a simple replication of the proof of [4, Lemma 3.3].

Recall that $\xi_i^{(k)} \stackrel{\text{def}}{=} \mathbb{T}^{(k)}(\nabla_i f_\mu(x^{(k)})) \in [-1, 1]$ is the truncated gradient at step $k$, and satisfies that $\xi_i^{(k)} = \nabla_i f_\mu(x^{(k)})$ for all coordinates $i$ such that $\nabla_i f_\mu(x^{(k)}) \in [-1, 1] \setminus [-\varepsilon, \varepsilon]$. We can verify that our careful choice of $x^{(k)} \to x^{(k+1)}$ is in fact a mirror descent step on the truncated gradient:

**Claim B.1.**
$$x^{(k+1)} = \arg\min_{z \geq 0} \left\{ V_{x^{(k)}}(z) + \langle \alpha\xi^{(k)}, z - x^{(k)} \rangle \right\} \ . \tag{B.3}$$

*Proof.* This can be verified coordinate by coordinate, because the arg min function is over all possible $z \geq 0$, where this constraint does not impose any inter-coordinate constraint.

In other words, by substituting the definition of $V_{x^{(k)}}(z)$, we only need to verify that

$$x_i^{(k+1)} = \arg\min_{z_i \geq 0} \left\{ \left( z_i \log \frac{z_i}{x_i^{(k)}} + x_i^{(k)} - z_i \right) + \alpha\xi_i^{(k)} \cdot (z_i - x_i^{(k)}) \right\} \stackrel{\text{def}}{=} \arg\min_{z_i \geq 0} \{g(z_i)\} \ .$$

At this point, the univariate function $g(z_i)$ is convex and has a unique minimizer. Since the gradient $\frac{d}{dz_i} g(z_i) = \log \frac{z_i}{x_i^{(k)}} + \alpha\xi_i^{(k)}$, this unique minimizer is indeed $z_i = x_i^{(k)} \cdot e^{-\alpha\xi_i^{(k)}}$, finishing the proof of Claim B.1. $\qquad \square$

After confirming that our iterative step in `PosSDPSolver` is indeed a mirror descent step, it is not hard to deduce Lemma 5.2 based on the proof of the classical mirror descent analysis.

*Proof of Lemma 5.2.* We deduce the following sequence of inequalities:

$$\begin{aligned}
\langle \alpha \xi^{(k)}, x^{(k)} - u \rangle &= \langle \alpha \xi^{(k)}, x^{(k)} - x^{(k+1)} \rangle + \langle \alpha \xi^{(k)}, x^{(k+1)} - u \rangle \\
&\stackrel{\text{①}}{=} \langle \alpha \xi^{(k)}, x^{(k)} - x^{(k+1)} \rangle + \langle -\nabla V_{x^{(k)}}(x^{(k+1)}), x^{(k+1)} - u \rangle \\
&\stackrel{\text{②}}{=} \langle \alpha \xi^{(k)}, x^{(k)} - x^{(k+1)} \rangle + V_{x^{(k)}}(u) - V_{x^{(k+1)}}(u) - V_{x^{(k)}}(x^{(k+1)}) \\
&\stackrel{\text{③}}{\leq} \sum_i \left( \alpha \xi_i^{(k)} \cdot (x^{(k)} - x^{(k+1)}) - \frac{|x_i^{(k+1)} - x_i^{(k)}|^2}{2 \max\{x_i^{(k+1)}, x_i^{(k)}\}} \right) + \left( V_{x^{(k)}}(u) - V_{x^{(k+1)}}(u) \right) \\
&\stackrel{\text{④}}{\leq} \sum_i \frac{(\alpha^2 \xi_i^{(k)})^2 \cdot \max\{x_i^{(k+1)}, x_i^{(k)}\}}{2} + \left( V_{x^{(k)}}(u) - V_{x^{(k+1)}}(u) \right) \quad \text{(B.4)} \\
&\stackrel{\text{⑤}}{\leq} \frac{2}{3} \alpha^2 \mathbb{1}^T x^{(k)} + \left( V_{x^{(k)}}(u) - V_{x^{(k+1)}}(u) \right) \\
&\stackrel{\text{⑥}}{\leq} \alpha^2 \mathsf{OPT} + \left( V_{x^{(k)}}(u) - V_{x^{(k+1)}}(u) \right)
\end{aligned}$$

Here, ① is due to the minimality of $x^{(k+1)}$ in (B.3), which implies that $\nabla V_{x^{(k)}}(x^{(k+1)}) + \alpha \xi^{(k)} = 0$. ② is due to the triangle equality of Bregman divergence:

$$\begin{aligned}
\forall x, y \geq 0, \quad \langle -\nabla V_x(y), y - u \rangle &= \langle \nabla w(x) - \nabla w(y), y - u \rangle \\
&= (w(u) - w(x) - \langle \nabla w(x), u - x \rangle) - (w(u) - w(y) - \langle \nabla w(y), u - y \rangle) \\
&\quad - (w(y) - w(x) - \langle \nabla w(x), y - x \rangle) \\
&= V_x(u) - V_y(u) - V_x(y) \ .
\end{aligned}$$

③ is because $V_x(y) = \sum_i y_i \log \frac{y_i}{x_i} + x_i - y_i \geq \sum_i \frac{1}{2 \max\{x_i, y_i\}} |x_i - y_i|^2$. ④ is by Cauchy-Schwarz. ⑤ is because we have $x_i^{(k+1)} \leq \frac{4}{3} x_i^{(k)}$ owing to Fact 5.5. ⑥ is because we have $\mathbb{1}^T x^{(k)} \leq \frac{3}{2} \mathsf{OPT}$ owing to Proposition 4.2.d (and $f_\mu(x^{(k)}) \leq 0$ from Lemma 5.2). □

### B.4 Proof of Theorem 5.4

*Proof of Theorem 5.4.* We begin by telescoping the inequality in Lemma 5.3 for $k = 0, 1, \ldots, T-1$, and choosing $u = \widetilde{u} \stackrel{\text{def}}{=} (1 - \varepsilon/2) x^*$, which satisfies $\mathbb{1}^T u \leq \mathsf{OPT}$ by the definition of $x^*$:

$$\mathbf{E}\left[ \alpha \sum_{k=0}^{T-1} (f_\mu(x^{(k)}) - f_\mu(\widetilde{u})) \right] \leq 4(f_\mu(x^{(0)}) - \mathbf{E}[f_\mu(x^{(T)})]) + 2\left( V_{x^{(0)}}(\widetilde{u}) - \mathbf{E}[V_{x^{(T)}}(\widetilde{u})] \right) + \alpha T \cdot 3\varepsilon \mathsf{OPT} \ . \quad \text{(B.5)}$$

Above, the expectation is over the randomness of the entire algorithm. Notice that, the second term on the right hand side of (B.5) is upper bounded by

$$\begin{aligned}
V_{x^{(0)}}(\widetilde{u}) - \mathbf{E}[V_{x^{(T)}}(\widetilde{u})] &\leq V_{x^{(0)}}(\widetilde{u}) \leq \sum_i \widetilde{u}_i \log \frac{\widetilde{u}_i}{x_i^{(0)}} + x_i^{(0)} \leq \sum_i \widetilde{u}_i \log \frac{1/\|A_i\|_{\mathsf{spe}}}{(1 - \varepsilon/2)/n \|A_i\|_{\mathsf{spe}}} + \frac{1 - \varepsilon/2}{n \|A_i\|_{\mathsf{spe}}} \\
&\leq \mathbb{1}^T \widetilde{u} \cdot \log(2n) + 1 \leq 2\mathsf{OPT} \cdot \log(2n) \ . \quad \text{(B.6)}
\end{aligned}$$

Here, we have used the fact that $\widetilde{u}_i \leq \frac{1}{\|A_i\|_{\mathsf{spe}}}$ since $\widetilde{u}_i A_i \preceq I$.

From here, we want to prove that $\mathbf{E}[f_\mu(x^{(T)})] \leq -(1-5\varepsilon)\mathsf{OPT}$ by way of contradiction. Suppose not, that is, $\mathbf{E}[f_\mu(x^{(T)})] > -(1-5\varepsilon)\mathsf{OPT}$, we have $f_\mu(x^{(0)}) - \mathbf{E}[f_\mu(x^{(T)})] \leq 0 + (1-5\varepsilon)\mathsf{OPT} \leq \mathsf{OPT}$, giving an upper bound on the first term on the right hand side in (B.5). Substituting this and (B.6) to (B.5), and dividing $\alpha T$ on both sides, we get

$$\frac{1}{T}\sum_{k=0}^{T-1}(\mathbf{E}[f_\mu(x^{(k)})] - f_\mu(\widetilde{u})) \leq \frac{4}{\alpha T}(f_\mu(x^{(0)}) - \mathbf{E}[f_\mu(x^{(T)})]) + \frac{2}{\alpha T}\big(V_{x^{(0)}}(\widetilde{u}) - \mathbf{E}[V_{x^{(T)}}(\widetilde{u})]\big) + 3\varepsilon\mathsf{OPT}$$

$$\leq \frac{4\mathsf{OPT}}{\alpha T} + \frac{4\mathsf{OPT}\cdot\log(2n)}{\alpha T} + 3\varepsilon\mathsf{OPT} \ .$$

Finally, since we have chosen $T \geq \frac{8\log(2n)}{\alpha\varepsilon}$, the above right hand side is no greater than $4\varepsilon\mathsf{OPT}$. This, by an averaging argument, tells us the existence of some $k \in \{0, 1, \ldots, T-1\}$ with $\mathbf{E}[f_\mu(x^{(k)})] \leq f_\mu(\widetilde{u}) + 4\varepsilon\mathsf{OPT} \leq -(1-5\varepsilon)\mathsf{OPT}$ (where we have used $f_\mu(\widetilde{u}) \leq -(1-\varepsilon)\mathsf{OPT}$ from Proposition 4.2.b). However, it contradicts to the hypothesis that $\mathbf{E}[f_\mu(x^{(T)})] > -(1-5\varepsilon)\mathsf{OPT}$ because $f_\mu(x^{(k)}) \geq f_\mu(x^{(T)})$ according to Lemma 5.1. This finishes the proof that $\mathbf{E}[f_\mu(x^{(T)})] \leq -(1-5\varepsilon)\mathsf{OPT}$.

The fact that $\frac{x^{(T)}}{1+\varepsilon}$ provides a $(1-O(\varepsilon))$ approximate solution for the packing SDP is due to Proposition 4.2.e and Markov's inequality which states that $f_\mu(x^{(T)}) \leq -(1-O(\varepsilon))\mathsf{OPT}$ with at least constant probability. □

## C Missing Proofs for Section 6

The proof of Lemma 6.1 is completely analogous to its LP variant in [4]. We include it only for the sake of completeness.

**Lemma 6.1.** *For any $T \geq \frac{8}{\alpha\varepsilon} = \Omega(\frac{\log(nm/\varepsilon)}{\varepsilon^3})$, we have that $\mathbf{E}[\mathrm{Tr}(\overline{Y})] \leq (1+7\varepsilon)\mathsf{OPT}$.*

*Proof.* Telescoping Lemma 5.3 for $k = 0, 1, \ldots, T-1$ and $u = 0$, we have that

$$\frac{1}{T}\mathbf{E}\Big[\sum_{k=0}^{T-1}\langle\nabla f_\mu(x^{(k)}), x^{(k)}\rangle\Big] \leq \frac{4}{\alpha T}(f_\mu(x^{(0)}) - \mathbf{E}[f_\mu(x^{(T)})]) + \frac{2}{\alpha T}\big(V_{x^{(0)}}(0) - \mathbf{E}[V_{x^{(T)}}(0)]\big) + 2\varepsilon\mathsf{OPT}$$

$$\leq \frac{4}{\alpha T}(f_\mu(x^{(0)}) - \mathbf{E}[f_\mu(x^{(T)})]) + \frac{2}{\alpha T}V_{x^{(0)}}(0) + 2\varepsilon\mathsf{OPT}$$

$$\leq \frac{4}{\alpha T}(f_\mu(x^{(0)}) - \mathbf{E}[f_\mu(x^{(T)})]) + \frac{2}{\alpha T} + 2\varepsilon\mathsf{OPT} \ . \tag{C.1}$$

Above, the last inequality uses the fact that $V_{x^{(0)}}(0) = \mathbb{1}^T x^{(0)} \leq 1$.

We now respectively lower and upper bound the two sides of (C.1) as follows. One one hand, using the definition of gradient, the left hand side of (C.1) is lower bounded as

$$\langle\nabla f_\mu(x^{(k)}), x^{(k)}\rangle = \sum_{i\in[n]} x_i^{(k)} A_i \bullet e^{\frac{1}{\mu}\big(\sum_{i\in[n]} x_i^{(k)} A_i - I\big)} - \mathbb{1}^T x^{(k)}$$

$$\geq (1-\varepsilon)I \bullet e^{\frac{1}{\mu}\big(\sum_{i\in[n]} x_i^{(k)} A_i - I\big)} - \mathbb{1}^T x^{(k)} - m\cdot\big(\frac{\varepsilon}{nm}\big)^4$$

$$= (1-\varepsilon)\mathrm{Tr}(Y(x^{(k)})) - \mathbb{1}^T x^{(k)} - m\cdot\big(\frac{\varepsilon}{nm}\big)^4 \ .$$

Above, the (only) inequality is because if $B \stackrel{\text{def}}{=} \sum_{i\in[n]} x_i^{(k)} A_i$ has eigenvalues $\lambda_1, \ldots, \lambda_m \geq 0$, then $\sum_{i\in[n]} x_i^{(k)} A_i \bullet e^{\frac{1}{\mu}\big(\sum_{i\in[n]} x_i^{(k)} A_i - I\big)} = \sum_{j\in[m]} \lambda_j \cdot e^{(\lambda_j - 1)/\mu}$. However, if there are some $\lambda_j$ satisfying $\lambda_j < 1 - \varepsilon$, the corresponding term $e^{\frac{1}{\mu}(\lambda_j-1)} \leq e^{-\varepsilon/\mu} = (\frac{\varepsilon}{nm})^4$ is very small, and there are at most $m$ such small terms. As a result, one must have $\sum_{j \in [m]} \lambda_j \cdot e^{(\lambda_j-1)/\mu} \geq (1-\varepsilon) \sum_{j \in [m]} \cdot e^{(\lambda_j-1)/\mu} - m \cdot (\frac{\varepsilon}{nm})^4 = (1-\varepsilon)I \bullet e^{\frac{1}{\mu}\left(\sum_{i \in [n]} x_i^{(k)} A_i - I\right)} - m \cdot (\frac{\varepsilon}{nm})^4$.

On the other hand, since $x_i^{(T)} A_i \preceq (1+\varepsilon)I$ by Proposition 4.2.d, we must have $\mathbb{1}^T x^{(T)} \leq (1+\varepsilon)\mathsf{OPT}$ by the definition of $\mathsf{OPT}$, and thus $f_\mu(x^{(T)}) \geq 0 - (1+\varepsilon)\mathsf{OPT}$. This gives an upper bound on the right hand side of (C.1) that is $\frac{4(1+\varepsilon)}{\alpha T}\mathsf{OPT} + \frac{2}{\alpha T} + 2\varepsilon\mathsf{OPT} \leq 3\varepsilon\mathsf{OPT}$, due to our choice of $T \geq \frac{8}{\alpha\varepsilon}$.

Together, we deduce from (C.1) that

$$(1-\varepsilon)\frac{1}{T}\sum_{k=0}^{T-1} \mathbf{E}\Big[\mathrm{Tr}(Y(x^{(k)})) - \mathbb{1}^T x^{(k)}\Big] - m \cdot (\frac{\varepsilon}{nm})^4 \leq 3\varepsilon\mathsf{OPT}$$

$$\implies \mathbf{E}[\mathrm{Tr}(\overline{Y})] = \mathrm{Tr}\mathbf{E}\Big[\frac{1}{T}\sum_k Y(x^{(k)})\Big] \leq \frac{1}{T}\sum_k \mathbf{E}[\mathbb{1}^T x^{(k)}] + 4\varepsilon\mathsf{OPT} \leq (1+\varepsilon)\mathsf{OPT} + 4\varepsilon\mathsf{OPT} ,$$

where the last inequality is from $\mathbb{1}^T x^{(k)} \leq (1+\varepsilon)\mathsf{OPT}$ for each $k$ (see Proposition 4.2.d). □

As mentioned earlier, our proof for Lemma 6.2 below is fundamentally different from its much weaker version in [4].

**Lemma 6.2.** *For any* $T \geq \frac{8}{\alpha\varepsilon}$, *with probability at least* $1 - \frac{\varepsilon}{100}$ *we have* $A_i \bullet \overline{Y} \geq 1 - 2\varepsilon$ *for all* $i \in [n]$.

*Proof.* For each iteration $k = 0, \ldots, T-1$ and coordinate $i \in [n]$, we denote by

- $\gamma_i^{(k)} \stackrel{\text{def}}{=} \mathbb{T}^{(k)}(\nabla_i f_\mu(x^{(k)})) \in [-1, 1]$ the actual truncated gradient, and
- $\xi_i^{(k)} \stackrel{\text{def}}{=} \frac{1}{2}\big(\mathbb{T}_-(\nabla_i f_\mu(x^{(k)})) + \mathbb{T}_+(\nabla_i f_\mu(x^{(k)}))\big) \in [-1/2, 1/2]$ the expected truncated gradient.

It is easy to verify that $\mathbf{E}[\gamma^{(k)}] = \xi^{(k)}$, where the expectation is over the random choice of $\mathbb{T}^{(k)}$. In addition, since $\nabla_i f_\mu(x^{(k)}) = 2\xi_i^{(k)}$ whenever $\nabla_i f_\mu(x^{(k)}) \in [-1, 1] \setminus [-\varepsilon, \varepsilon]$ owing to the definition of the thresholding functions, we automatically have

$$\nabla_i f_\mu(x^{(k)}) \geq 2\xi_i^{(k)} - \varepsilon .$$

In the first step, recalling that $x_i^{(T)} = x_i^{(0)} \cdot e^{-\alpha \sum_{k=0}^{T-1} \gamma_i^{(k)}}$ by the definition of our update rule (Line 8 of PosSDPSolver), and recalling that $x_i^{(T)} A_i \preceq (1+\varepsilon)I \prec 1.5I$ due to Proposition 4.2.d which implies $x_i^{(T)} \leq \frac{1.5}{\|A_i\|_{\mathsf{spe}}}$, we automatically have that for every $i \in [n]$, independent of the randomness of the algorithm, it always satisfies that

$$\frac{1}{T}\sum_{k=0}^{T-1} \gamma_i^{(k)} \geq -\frac{\log(1.5/(\|A_i\|_{\mathsf{spe}} \cdot x_i^{(0)}))}{\alpha T} \geq \frac{-\log(2n)}{\alpha T} \geq -\frac{\varepsilon}{8} .$$

Above, the second inequality is due to our choice of $x^{(0)}$, and the third inequality is due to our choice of $T$. Next, define $Z_{k,i} \stackrel{\text{def}}{=} \sum_{j=0}^{k-1}(\gamma_i^{(k)} - \xi_i^{(k)})$, we have that $\{Z_{k,i}\}_{k=1}^{T}$ is a martingale, satisfying that $\mathbf{E}[Z_{k,i}|Z_{1,i}, \ldots, Z_{k-1,i}] = Z_{k-1,i}$ and $|Z_{k,i} - Z_{k-1,i}| \leq 1/2$. By the Azuma-Hoeffding inequality, we have

$$\Pr\Big[\frac{1}{T}\sum_{k=0}^{T-1}(\xi_i^{(k)} - \gamma_i^{(k)}) < -\frac{\varepsilon}{4}\Big] = \Pr\Big[\frac{Z_{T,i}}{T} > \frac{\varepsilon}{4}\Big] \leq e^{\frac{-\varepsilon^2 T}{8}} \leq \frac{\varepsilon}{100n} .$$

By a union bound, with probability at least $1 - \varepsilon/100$, for every $i \in [n]$,

$$\frac{1}{T}\sum_{k=0}^{T-1} \nabla_i f_\mu(x^{(k)}) \geq \frac{1}{T}\sum_{k=0}^{T-1} 2\xi_i^{(k)} - \varepsilon = 2\frac{1}{T}\sum_{k=0}^{T-1}(\xi_i^{(k)} - \gamma_i^{(k)}) + 2\frac{1}{T}\sum_{k=0}^{T-1}\gamma_i^{(k)} - \varepsilon \geq 2 \cdot (-\frac{\varepsilon}{4}) - \frac{\varepsilon}{4} - \varepsilon > -2\varepsilon \enspace .$$

In other words, with probability at least $1 - \varepsilon/100$, for every $i \in [n]$,

$$A_i \bullet \overline{Y} - 1 = \frac{1}{T}\sum_{k=0}^{T-1}(A_i \bullet Y(x^{(k)}) - 1) = \frac{1}{T}\sum_{k=0}^{T-1} \nabla_i f_\mu(x^{(k)}) \geq -2\varepsilon \enspace . \qquad \square$$